\documentclass[conference]{IEEEtran}
\IEEEoverridecommandlockouts

\usepackage[table]{xcolor}
\usepackage{subcaption}
\usepackage{xspace}
\usepackage{tikz}
\usetikzlibrary{arrows.meta, shapes}
\usetikzlibrary {shadows}
\usetikzlibrary{fit, backgrounds, calc}
\usetikzlibrary{fadings}
\usetikzlibrary{decorations.pathmorphing}

\usepackage{subfig}
\usepackage{cite}
\usepackage{amsmath,amssymb,amsfonts}
\usepackage{algorithm}
\usepackage{algorithmicx}
\usepackage[noend]{algpseudocode}
\algrenewcommand\algorithmiccomment[2][\itshape]{{#1\hfill\(\triangleright\)
    #2}}
\algrenewcommand{\algorithmicrequire}{\textbf{Input:}}
\algrenewcommand{\algorithmicensure}{\textbf{Output:}}
\usepackage{graphicx}
\usepackage{textcomp}
\usepackage{fancyvrb}
\usepackage{listings}
\usepackage{xcolor}
\usepackage{float}
\usepackage{todonotes}
\usepackage{booktabs}
\usepackage{multirow}
\usepackage{tcolorbox}
\usepackage{enumitem}

\def\BibTeX{{\rm B\kern-.05em{\sc i\kern-.025em b}\kern-.08em
    T\kern-.1667em\lower.7ex\hbox{E}\kern-.125emX}}
    
\usepackage{etoolbox}
\makeatletter
\patchcmd{\@begintheorem}{\textit}{\textbf}{}{}
\makeatother

\usepackage{fontawesome5}
\usepackage{adjustbox}

\usepackage{bold-extra} 

\newcommand{\inlineheadingbf}[1]{\medskip\noindent{\bfseries #1.}}
\newcommand{\inlineheadingit}[1]{\medskip\noindent{\it #1.}}

\newcommand{\llbox}[1]{%
  \begin{tcolorbox}[
    width=\columnwidth,
    colback=black!6,
    colframe=black!6,
    boxrule=0pt,
    arc=2pt,
    top=1.5mm, bottom=1.5mm, left=2mm, right=1.5mm
  ]
  #1
  \end{tcolorbox}%
}

\newcommand{\desc}{d}
\newcommand{\interpretations}{\mathbb{I}}
\newcommand{\LLM}{\mathsf{LLM}}
\newcommand{\support}{\mathsf{supp}_{\equiv}}
\newcommand{\tests}{\mathcal{T}}

\usepackage{cleveref}
\crefname{figure}{Fig.}{Fig.}
\Crefname{figure}{Figure}{Figures}
\crefname{definition}{Def.}{Defs.}
\Crefname{definition}{Definition}{Definitions}
\crefname{table}{Tab.}{Tab.}
\Crefname{table}{Table}{Tables}
\crefname{equation}{Eq.}{Eq.}
\Crefname{equation}{Equation}{Equations}
\crefname{algorithm}{Alg.}{Alg.}
\Crefname{algorithm}{Algorithm}{Algorithms}
\crefname{section}{Sec.}{Sec.}
\Crefname{section}{Section}{Sections}

\definecolor{clusterfill}{RGB}{225, 236, 250}
\definecolor{clusterline}{RGB}{ 24,  95, 165}
\definecolor{grayfill}{RGB}{238, 238, 238}
\definecolor{grayline}{RGB}{170, 170, 170}
\definecolor{boxfill}{RGB}{245, 245, 245}
\definecolor{promptorange}{RGB}{230,132,0}
\definecolor{abstgreen}{RGB}{96,160,74}
\definecolor{badred}{RGB}{220,0,0}
\definecolor{promptbg}{RGB}{235,235,235}
\definecolor{goodbg}{RGB}{213,231,202}
\definecolor{badbg}{RGB}{243,210,210}
\definecolor{icongray}{RGB}{95,95,95}
\definecolor{charcoal}{rgb}{0.21, 0.27, 0.31}

\newcommand{\href}[2]{{\em #2}}

\def\BibTeX{{\rm B\kern-.05em{\sc i\kern-.025em b}\kern-.08em
    T\kern-.1667em\lower.7ex\hbox{E}\kern-.125emX}}
\begin{document}

\title{Underspecification does {\em not} imply Incoherence: The Risks of Semantic Collapse in Coding Models}

\author{
\IEEEauthorblockN{Cedric Richter}
\IEEEauthorblockA{
\textit{SnT, University of Luxembourg}\\
Luxembourg \\
cedric.richter@uni.lu}
\and
\IEEEauthorblockN{Mike Papadakis}
\IEEEauthorblockA{
\textit{SnT, University of Luxembourg}\\
Luxembourg \\
michail.papadakis@uni.lu}
}

\maketitle

\begin{abstract}
Large Language Models (LLMs) have become increasingly effective at generating code when task descriptions are clear and precise. Yet, in practice, user-provided task descriptions are often ambiguous, incomplete, or contradictory, leaving critical aspects of the intended program behavior underspecified. In such cases, multiple behaviorally distinct interpretations may satisfy the description equally well, yet semantically differ in ways that matter/affect the user intent. A natural expectation, often assumed by researchers, is that prompt underspecification manifests as {\em incoherence}: When asked multiple times, an LLM produces multiple semantically distinct implementations reflecting the ambiguity of the task description. In this paper, we challenge this assumption. We find that LLMs frequently collapse onto a single incorrect interpretation of the task description, consistently generating coherent but behaviorally misaligned code. We term this failure mode \emph{detrimental semantic collapse} and find that it affects over 10\% of tasks in MBPP, 3\% in HumanEval, and 32\% of LiveCodeBench, all benchmarks assumed to be well-specified. By deliberately injecting underspecification issues in the benchmark prompts, the rate rises to over 5 times, exposing a fundamental blind spot in disambiguation and correctness estimation techniques that rely on incoherence as a proxy for prompt underspecification.\\
\end{abstract}


\section{Introduction}
When a developer is asked to implement a function that retrieves the maximum value from a list, the task seems straightforward. But what should be returned if the list contains heterogeneous elements such as integers, floating-point numbers, or strings? The maximum integer, the largest number, or the lexicographically largest string? The task description does not say. 
Software developers are frequently confronted with task descriptions that are ambiguous, incomplete, or contradictory, leaving critical aspects of the intended behavior unspecified~\cite{DBLP:conf/birthday/GervasiFZS19, DBLP:conf/re/MavinWHN09, DBLP:journals/pacmse/Mu00YZWL024, DBLP:conf/kbse/JiaMYSM25}. In these cases, the developer has to disambiguate between multiple behaviorally distinct interpretations by asking clarifying questions, exploring alternatives, or committing to a single interpretation.

Today, software development has been largely transformed by large language models (LLMs), which increasingly automate tasks that traditionally require significant manual effort~\cite{DBLP:journals/ese/ZhengNZCCGWW25}. The most prominent such application is code generation~\cite{DBLP:journals/corr/abs-2107-03374, DBLP:journals/corr/abs-2203-07814, DBLP:conf/emnlp/ShiFGZW22}: given a natural language task description, an LLM can produce a working implementation. This raises a critical question: when the task description is underspecified, as is frequently the case in practice, how does the LLM behave?

A natural expectation is that underspecification leads to \emph{incoherence}: When repeatedly asked with the same underspecified task description, an LLM should produce multiple semantically distinct implementations reflecting the ambiguity of the task. This incoherence can be detected by clustering LLM-generated programs into equivalence classes based on their execution behavior. If multiple clusters emerge, the task description is deemed underspecified~\cite{DBLP:journals/pacmse/Mu00YZWL024, DBLP:conf/kbse/JiaMYSM25}.

This assumption has motivated several studies. Mu et al.~\cite{DBLP:journals/pacmse/Mu00YZWL024} use semantic clustering to identify
underspecified task descriptions and generate clarifying questions. Jia et al.~\cite{DBLP:conf/kbse/JiaMYSM25} cluster generated programs to automatically repair ambiguous task descriptions. More recent work~\cite{DBLP:journals/corr/abs-2502-11620, DBLP:conf/aaai/ValentinMSB26} has also started to use semantic clustering as a proxy for program incorrectness, treating incoherent outputs as evidence of incorrectness. All of these studies focus on the implications of LLM incoherence, leaving open a critical question: \emph{Does underspecification imply incoherence?} 

A positive answer would reinforce the applicability of the related studies and motivate further efforts to improve code correctness and, more broadly, the reliability of automatic coding systems. Conversely, a negative answer would highlight a significant limitation of existing methods and further emphasize both the importance and the difficulty of mitigating the issues arising from the prompt underspecification.

In this work, we answer the above question by systematically studying whether prompt underspecification leads to incoherence of coding models. We evaluate three state-of-the-art coding LLMs, namely Claude Sonnet 4.5~\cite{anthropic2025claude}, GPT-4.1-mini~\cite{DBLP:journals/corr/abs-2303-08774}, and Qwen3~\cite{DBLP:journals/corr/abs-2505-09388}, on systematically underspecified variants of three widely used coding benchmarks: MBPP~\cite{DBLP:journals/corr/abs-2108-07732}, HumanEval~\cite{DBLP:journals/corr/abs-2107-03374}, and LiveCodeBench~\cite{DBLP:conf/iclr/JainHGLYZWSSS25}.

\begin{figure*}[t]
\vspace{1.0em}
\centering
\begin{subfigure}{0.26\linewidth}
    \resizebox{\linewidth}{!}{
\begin{tikzpicture}[
    x=1cm,y=1cm,
    every node/.style={inner sep=0pt, outer sep=0pt},
    title/.style={font=\bfseries\fontsize{18}{20}\selectfont},
    sectiontitle/.style={font=\bfseries\fontsize{16}{18}\selectfont},
    bubbletext/.style={font=\ttfamily\fontsize{15}{17}\selectfont, align=center},
    iconlabel/.style={font=\fontsize{22}{24}\selectfont},
]


\node[sectiontitle, text=blue!60!black] (user-title) at (0, 0) {Underspecified Prompt - MBPP/294};

\node[iconlabel, text=icongray, below left=0.2cm and 0cm of user-title] (user) {\faUser};

\node[bubbletext, fill=promptbg, above right=-2.6cm and 0.2cm of user, text width=8.1cm, rounded corners = 0.22cm, inner sep=0.5cm, align=left]  (prompt) {%
Write a function to find the \colorbox{badbg}{maximum value} in a given heterogeneous list.
};

\node[sectiontitle, text=black, below right=0.5cm and -6cm of prompt] (title-incorrect) {Consistent Model Response};

\node[iconlabel, text=icongray, below right=0.2cm and 0cm of title-incorrect] (robot) {\faRobot};

\node[bubbletext, anchor=center, text width=9cm, rounded corners=0.24cm,  fill=promptbg,  above left=-4.4cm and 0.2cm of robot, align=left ] (answer) {%
\begin{minipage}{\linewidth}
\adjustbox{max width=1.2\linewidth}{
    \begin{lstlisting}[numbers=none, breaklines=false, escapechar=!]
def max_val(lst):
  numeric_values = [
  	item for item in lst 
  	if isinstance(item, 
		!{\color{red}(int, float)}!)
  ]
  ...
    \end{lstlisting}
  }
 \end{minipage}
};

\node[bubbletext, anchor=center, text width=8cm, rounded corners=0.24cm,  fill=badbg,  below left=-0.5cm and -8.7cm of answer, inner sep=0.2cm, align=left ] (verdict) {%
\faThumbsDown[regular] ~ maximum {\em integer} value
};

\node [below=0.7cm of verdict]{};

\end{tikzpicture}}
    \caption{Example}\label{fig:mbpp}
\end{subfigure}
\hfill
\begin{subfigure}{0.36\linewidth}
     \resizebox{\linewidth}{!}{
\begin{tikzpicture}[
    description/.style = {draw, rounded corners=2pt, fill=boxfill,
                          minimum width=1.4cm, minimum height=1.0cm,
                          align=center, font=\small},
    llm/.style         = {draw, rounded corners=2pt, fill=boxfill,
                          minimum width=1.2cm, minimum height=0.7cm,
                          font=\small\bfseries},
    cluster/.style     = {draw=clusterline, fill=clusterfill,
                          line width=0.8pt, circle},
    graycluster/.style = {draw=grayline, fill=grayfill,
                          line width=0.8pt, circle},
    flow/.style        = {-{Stealth}, line width=0.8pt},
    rowlabel/.style    = {font=\bfseries, anchor=west},
    caption/.style     = {font=\small, align=center},
  ]

  \newcommand{\dotat}[4][clusterline]{\fill[#1] ([shift={(#3,#4)}]#2.center) circle (1.2pt);}
  \newcommand{\graydotat}[3]{\fill[grayline] ([shift={(#2,#3)}]#1.center) circle (1.2pt);}

  \node[rowlabel] at (-0.8, 3.8) {Expectation};
  \draw[thick, dashed] (1.6, 3.8) -- (11.0, 3.8);

  \node[fill=promptbg, rounded corners = 0.22cm,  text width=2cm, minimum width=1.4cm, minimum height=1.0cm,
                          align=center, font=\small] (descE) at (0.6, 2.4) {
  {\scalebox{3.5}{\faFile*[regular]}}
  };
  \node [below=0cm of descE, text width=2cm, align=center, text=blue!60!black, font=\bfseries] (title){Underspecified Prompt};
  \node[llm]        (llmE)  at (3.0, 2.4) {LLM};
  \draw[flow] (descE) -- (llmE);
  \draw[flow] (llmE) -- ++(1.1,0);

  \node[cluster, minimum size=1.6cm] (e1) at (5.6, 2.4) {};
  \node[cluster, minimum size=1.4cm, fill=orange!30, draw=orange] (e2) at (7.6, 2.4) {};
  \node[cluster, minimum size=1.3cm, fill=green!80!black!20, draw=green!80!black] (e3) at (9.4, 2.4) {};
  \node (edots) at (10.7, 2.4) {$\cdots$};

  \dotat{e1}{-0.2}{0.2} \dotat{e1}{0.25}{0.25} \dotat{e1}{0.0}{-0.1}
  \dotat{e1}{-0.25}{-0.25} \dotat{e1}{0.3}{-0.2}
  \dotat[orange]{e2}{-0.15}{0.15} \dotat[orange]{e2}{0.2}{-0.1} \dotat[orange]{e2}{-0.1}{-0.2}
  \dotat[green!80!black]{e3}{-0.1}{0.1} \dotat[green!80!black]{e3}{0.15}{-0.05}

  \node[caption, below=0.15cm of e1] {dominant\\interpretation};

  \node[rowlabel] at (-0.8, 0.2) {Semantic Collapse};
  \draw[thick, dashed] (2.4, 0.2) -- (11.0, 0.2);

  \node[fill=promptbg, rounded corners = 0.22cm,  text width=2cm, minimum width=1.4cm, minimum height=1.0cm,
                          align=center, font=\small] (descR) at (0.6, -1.2)  {
  {\scalebox{3.5}{\faFile*[regular]}}
  };
  \node [below=0cm of descR, text width=2cm, align=center, text=blue!60!black, font=\bfseries] (title){Underspecified Prompt};
  \node[llm]        (llmR)  at (3.0, -1.2) {LLM};
  \draw[flow] (descR) -- (llmR);
  \draw[flow] (llmR) -- ++(1.1,0);

  \node[cluster, minimum size=2.0cm]    (r1) at (5.8, -1.2) {};
  \node[graycluster, minimum size=0.7cm](r2) at (8.0, -1.2) {};
  \node[graycluster, minimum size=0.7cm](r3) at (9.4, -1.2) {};

  \dotat{r1}{-0.3}{0.3}  \dotat{r1}{0.1}{0.35} \dotat{r1}{0.35}{0.1}
  \dotat{r1}{-0.35}{-0.1}\dotat{r1}{0.0}{0.0}  \dotat{r1}{0.3}{-0.3}
  \dotat{r1}{-0.1}{-0.35}\dotat{r1}{0.2}{-0.1} \dotat{r1}{-0.25}{0.05}

  \node[caption, below=0.15cm of r1] {dominant\\interpretation};

\end{tikzpicture}}
    \caption{Semantic Clustering}\label{fig:clustering}
\end{subfigure}
\hfill
\begin{subfigure}{0.36\linewidth}
    \includegraphics[width=\linewidth]{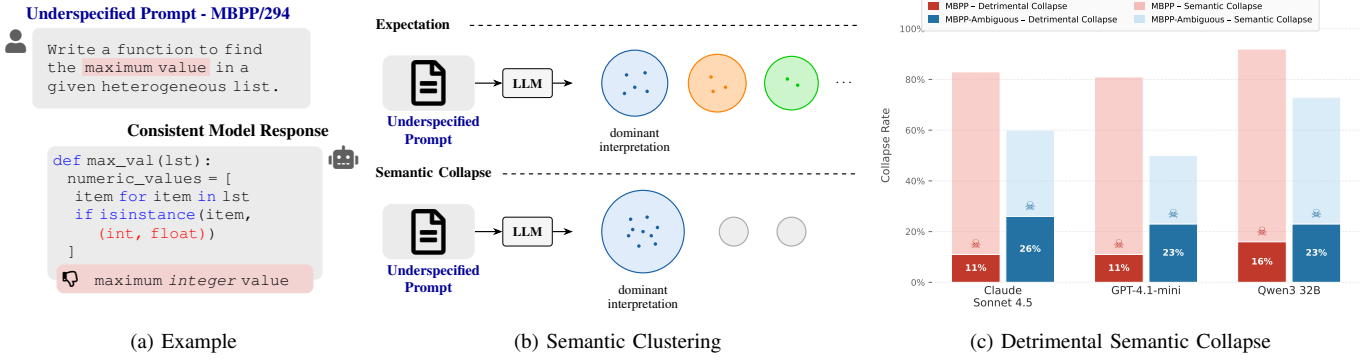}
    \caption{Detrimental Semantic Collapse}\label{fig:collapse}
\end{subfigure}
\vspace{1.0em}
\caption{\textbf{Motivating example.} (a) Real-world programming tasks are often underspecified. (b) Semantic clustering approaches assume that underspecified prompts lead LLMs in producing multiple semantically different interpretations when prompted multiple times (with the same prompt/description). We observe that such a phenomenon happens, but more often it does not as LLMs collapse to a single interpretation. (c) Collapse is often detrimental, producing a single incorrect interpretation breaking the key assumption made by semantic clustering methods. For example underspecification manifested in form of ambiguous descriptions, increases the risk for detrimental collapse. }
\label{fig:motivation}
\end{figure*}

Contrary to the prevailing assumption, we find that prompt underspecification does not imply incoherence. Instead, LLMs frequently collapse to a single incorrect interpretation, while remaining coherent across repeated generations. We term this failure mode \emph{detrimental semantic collapse}, as the LLM silently commits to a wrong interpretation, providing neither a correct solution nor any indication that the task description was underspecified.

We further show that detrimental semantic collapse is prevalent even on benchmarks widely regarded as well-specified, affecting over 3\% of HumanEval, 11\% of MBPP, and 32\% of LiveCodeBench tasks. Under controlled underspecification, collapse rates can increase by a factor of up to 5.5x on individual benchmarks, suggesting that current LLMs are substantially less capable of surfacing ambiguity than assumed.

\inlineheadingbf{Motivating example} \Cref{fig:mbpp} shows the task description of MBPP/294, a benchmark task of the popular MBPP~\cite{DBLP:journals/corr/abs-2108-07732} coding benchmark, and a response from Claude Sonnet 4.5. The LLM is tasked to implement a function to retrieve the maximum value in a heterogeneous list. Interestingly, the description does not specify what the maximum value in a list of integers, floats, and strings is.  The LLMs consistently assumes across multiple trials that the function should return the maximum {\em number} (floats and integers) in a list, yet the developer tests reveal that the actual intention was to only retrieve the maximum {\em integer} value. Unfortunately, the consequence of this underspecification issue is that the LLM collapses to a single incorrect interpretation, even though in practice multiple feasible interpretations exist. 

MBPP/294 is one of several benchmark tasks that we manually identified as ambiguous\footnote{Examples of Ambiguity: MBPP/294, MBPP/102, MBPP/410, MBPP/576}
, incomplete\footnote{Incompletness: MBPP/7, MBPP/137, MBPP/244, MBPP/261, MBPP/278}
, or contradictory\footnote{Contradictions: MBPP/459, MBPP/638, MBPP/639}. The thing is, as we show in this paper, that repeatedly sampling a response from an LLM produces the same incorrect interpretation each time. 

This is particularly problematic for clustering-based approaches (\Cref{fig:clustering}) that assume that underspecification yields distinct semantic clusters representing different interpretations, e.g., one cluster supporting integer inputs only and another supporting all numeric types. Under detrimental semantic collapse, clustering-based approaches fail silently, failures evade detection, as the incorrect implementation is consistent across generations, leaving users unaware of the underlying misalignment. Our experiments show that detrimental semantic collapse appears frequently across models, and is further amplified by underspecification (as shown for MBPP in \Cref{fig:collapse}).

\inlineheadingbf{Contributions} Overall, our work makes the following three key contributions:

\begin{itemize}

\item We identify the \emph{detrimental semantic collapse}, a previously overlooked failure mode in which coding LLMs consistently commit to a single behaviorally misaligned interpretation of an underspecified task, providing neither a correct solution nor any signal of incorrectness or underspecification.

\item We show that prompt underspecification frequently does not induce incoherence in coding LLMs, challenging a foundational assumption of semantic clustering-based approaches.

\item Through experiments on three state-of-the-art coding models and systematically underspecified variants of MBPP, HumanEval, and LiveCodeBench, we demonstrate that detrimental collapse rates can rise from 10\% on original benchmark tasks (MBPP) to over 50\% on systematically underspecified variants.

\end{itemize}

More broadly, our findings expose a fundamental limitation of approaches that use incoherence as a proxy for prompt ambiguity or program incorrectness: coherent model behavior cannot be interpreted as evidence of correct task understanding. In practice, this means that under-specified prompts cannot be reliably detected using the LLMs' interpretations, thereby limiting the confidence one can place in generated code without performing any independent testing. 

These results can be seen as a `call to arms' for the research community to develop more principled methods for evaluating prompts and model understanding, including formal specifications, robustness benchmarks, and verification frameworks that go beyond surface-level coherence.
\section{Related Work}
\begin{figure}[t]
    \centering
    \vspace{0.5em}
    \includegraphics[width=\columnwidth]{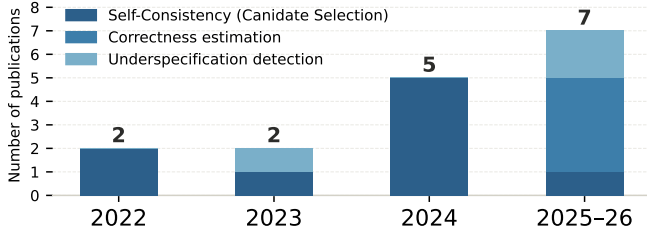}
        \vspace{0.5em}
    \caption{Number of publications published in top-tier SE and AI conferences that rely on the semantic clustering of LLM outputs as a proxy signal, grouped by year and per research topic/purpose, i.e., solution selection, correctness estimation, and prompt quality detection. The 2025--26 bar combines both years as 2026 is ongoing.}
                 \vspace{0.5em}
    \label{fig:related-work-trend}
\end{figure}

We observe an increasing trend of methods that rely on {\em semantic incoherence} in the scientific literature, results of a mini-collection survey, performed in an ad-hoc procedure that checks the titles of papers published in SE (FSE, ASE, and ISSTA) and AI  (NeurIPS, ICLR, ACL, EMNLP, and AAAI) conferences, over the past 4 years is shown in \Cref{fig:related-work-trend}. The topics of these studies are related to (1) candidate selection, (2) estimating correctness, and (3) identifying underspecification, with topics (2) and (3) gaining popularity over the years. 

\inlineheadingit{Self-consistent Candidate Selection} Self-consistency~\cite{DBLP:conf/iclr/0002WSLCNCZ23} is an effective candidate selection method that only reports the answer that occurs most often across sampled LLM responses. Self-consistency has been adapted in many forms to the coding domain: MBR-Exec~\cite{DBLP:conf/emnlp/ShiFGZW22} selects a program that is most consistent with all other sampled programs when executed on a small set of test inputs. DOCE~\cite{DBLP:journals/corr/abs-2408-13745} combines MBR-Exec's selection strategy with self-debugging. AlphaCode~\cite{DBLP:journals/corr/abs-2203-07814} clusters generated programs by executing them on generated test inputs and picks candidate solutions from the largest clusters. 

CodeT~\cite{DBLP:conf/iclr/ChenZNZLLC23} samples multiple programs and test inputs and selects candidate programs that have the highest code-test consistency. MPSC~\cite{DBLP:conf/acl/HuangL0D24} aligns generates programs, formal specifications, and test cases while selecting programs that have the strongest execution alignment with tests and specifications. SRank~\cite{DBLP:conf/acl/ToNB24} clusters and ranks code candidates by their execution similarity. LLMCodeChoice~\cite{DBLP:conf/issta/FanRMR24} clusters programs by execution, but leaves the selection to an LLM judge. 

FunCoder~\cite{DBLP:conf/nips/ChenTCCW0024} uses execution similarity to select sub-routines in larger coding contexts. Functional Majority Voting~\cite{DBLP:journals/corr/abs-2604-15618} has recently been proposed as a way for execution-based clustering and test-time training. All these techniques share a common assumption: Mistakes in LLM-generated code produce divergent behavior on individual test inputs, while correct behavior is shared across a majority of implementations. 

Interestingly, we find that detrimental semantic collapse represents a {\em major bottleneck} of these methods: When the LLM's output collapses to a single semantically incorrect interpretation, self-consistency methods fail to identify any correct solution. 

\inlineheadingit{Correctness} There is a recent surge in interest of using semantic incoherence of LLMs as an oracle-less measure of incorrectness~\cite{DBLP:journals/corr/abs-2502-11620, DBLP:conf/aaai/ValentinMSB26, DBLP:journals/corr/abs-2506-11021, DBLP:journals/corr/abs-2603-29292, DBLP:journals/corr/abs-2604-15618}. Sharma and David~\cite{DBLP:journals/corr/abs-2502-11620} found that the count of semantic clusters correlates with code correctness, e.g. a lower number of clusters indicates a higher correctness likelihood. Valentin et al.~\cite{DBLP:conf/aaai/ValentinMSB26} established semantic incoherence as a lower bound on model error. Ravuri and Amarasinghe~\cite{DBLP:journals/corr/abs-2506-11021} further employ semantic incoherence to identify hallucination-induced errors. 

Based on these insights, recent methods started to use semantic coherence as a pseudo-label for program correctness~\cite{DBLP:journals/corr/abs-2603-29292, DBLP:journals/corr/abs-2604-15618}. Our experiments indicate a {\em major  blind spot} of these methods: While semantic incoherence is an effective way to identify potential model error, its absence is not a reliable indicator of model correctness. 

\inlineheadingit{Underspecification} While semantic incoherence has traditionally been interpreted as a symptom of model error, recent work has evaluated semantic incoherence as a signal of prompt underspecification, using variance across generations to identify and clarify ambiguous or incomplete task descriptions~\cite{DBLP:journals/pacmse/Mu00YZWL024, DBLP:journals/corr/abs-2504-16331, DBLP:conf/kbse/JiaMYSM25}. 

ClarifyGPT~\cite{DBLP:journals/pacmse/Mu00YZWL024} uses semantic clustering to identify underspecified prompts and exploits semantic differences between candidates in different clusters to ask clarifying questions. ClarifyCoder~\cite{DBLP:journals/corr/abs-2504-16331} employs the clarification strategy of ClarifyGPT to generate training examples and train language models to ask clarifying questions. 

SpecFix~\cite{DBLP:conf/kbse/JiaMYSM25} exploits semantic differences of candidates in different clusters to localize and repair problems in task descriptions. While often effective in practice, these technique rely on the assumption that underspecified task descriptions produce divergent semantic clusters. 

Unfortunately, we observe that specifically prompt underspecification increases the risk of detrimental semantic collapse, rendering these methods ineffective on a large fraction of underspecified prompts.

\section{Objectives and Research Questions}
We aim to assess the behavior of large language models on the task of code generation from task descriptions especially when these are \emph{underspecified}. Formally, given a textual task description $\desc$ of programming task, the goal of the task is to generate the intended target program $P^*$. The task description $\desc$ is underspecified by admitting multiple feasible interpretations:
\begin{equation}
\interpretations(\desc) = \{ P_1, P_2, \dots \},
\end{equation}
where $P^* \in \interpretations(\desc)$, i.e. the intended program is one of many feasible interpretations. 

We are specifically interested in whether large language models reflect the ambiguity of underspecified task descriptions through semantically diverse responses, or whether they collapse to a single interpretation, the key assumption made by clustering-based approaches, as we already discussed. 

\subsection{Research Questions}
We investigate the behavior of large language models when confronted with underspecified task descriptions by answering the following research questions (RQs):

\begin{description}
\item[\textbf{RQ1.}] \textbf{Do LLMs explore multiple interpretations of underspecified task descriptions?} A natural assumption is that LLMs produce multiple inconsistent interpretations when prompted with underspecified task descriptions. We investigate whether this is true by measuring whether underspecified task descriptions lead to semantically diverse responses.\\

\item[\textbf{RQ2.}] \textbf{How frequently does semantic collapse become detrimental?} While RQ1 examines whether underspecification increases inconsistency of LLM responses, inconsistency alone does not indicate correctness, nor does semantic collapse necessarily imply incorrectness. We measure the rate of detrimental collapse across both well-specified and underspecified task descriptions, investigating whether underspecification elevates the likelihood that a collapsed interpretation is incorrect despite being internally consistent.\\

\item[\textbf{RQ3.}] \textbf{What is the impact of detrimental semantic collapse on downstream techniques?} Methods that rely on semantic incoherence to detect prompt underspecification, such as clarification question generation, presuppose that underspecified prompts produce multiple distinct interpretations. We evaluate the extent to which detrimental collapse constitutes a blind spot for such techniques, causing them to silently return incorrect functionality without triggering clarification.\\
\end{description}

\subsection{Semantic Clustering and Collapse}

\inlineheadingbf{Semantic Clustering} Two programs $P$ and $P'$ are {\em semantically indistinguishable}  with respect to a set of test inputs $\tests$ ($P \equiv_{\tests} P'$) if:
\begin{equation}
\forall x \in \tests: P(x) = P'(x),
\end{equation}
We cluster sampled programs into equivalence classes $[P]_{\equiv_{\tests}}$, where $P' \in [P]_{\equiv_{\tests}}$ if $P'$ is semantically indistinguishable from $P$. Each equivalence class corresponds to a distinct interpretation of the task description $\desc$.  

\inlineheadingbf{Semantic Collapse}
Let $\support(\desc)$ be the set of equivalence classes\footnote{For brevity, we use equivalence classes and their representatives (programs) interchangeably.} obtained by sampling from $\LLM$. We define the support $\support(\desc)$ as {\em semantically collapsed} iff:
\begin{equation}
\left\vert \support(\desc) \right\vert = 1,
\end{equation}
i.e. repeatedly sampling from the $\LLM$ yields a single interpretation, despite $\left\vert \interpretations(\desc) \right\vert > 1$.

Semantic collapse is {\em benign} when $\support(\desc) = \{P^*\}$, i.e. the LLM consistently produces the same interpretation, and {\em detrimental} when $\support(\desc) = \{P' \}$ with $P' \neq P^*$, i.e. the LLM converges on a single incorrect interpretation. 

Critically, benign and detrimental collapse are observably indistinguishable without access to an oracle: The distribution of LLM responses provides no indication that the task description was underspecified.

\subsection{Measuring Semantic Collapse under Finite Budget}\label{sec:measuring}

\begin{algorithm}[t]
\caption{Sequential Clustering for Collapse Detection}
\label{alg:collapse}
\begin{algorithmic}[1]
\Require Task description $d$, language model $\LLM$, a set of test inputs $\tests$, budget $k$
\Ensure Set of cluster representatives $R$

\State $R \gets \emptyset$ \Comment{Set of cluster representatives}
\For{$i \gets 1$ \textbf{to} $k$}
    \State $P_i \sim \LLM(d)$ \Comment{Sample program from LLM}
    \State $\textit{matched} \gets \textbf{false}$
    \For{$P \in R$}
        \If{$\forall x \in \tests: P_i (x) = P(x)$}
            \State $\textit{matched} \gets \textbf{true}$
            \State \textbf{break}
        \EndIf
    \EndFor
    \If{\textbf{not} $\textit{matched}$}
        \State $R \gets R \cup \{P_i\}$
        \If{$|R| > 1$}
            \State \Return $R$ \Comment{Collapse refuted}
        \EndIf
    \EndIf
\EndFor
\State \Return $R$ \Comment{Collapse confirmed: $|R| = 1$}
\end{algorithmic}
\end{algorithm}

\inlineheadingbf{Finite approximation of $\support(\desc)$} While semantic collapse cannot be confirmed with certainty from a finite set of samples, we can derive an efficient detection algorithm that provides PAC-style guarantees~\cite{DBLP:conf/aaai/ValentinMSB26}. Assume we have observed independent samples $P_1, P_2, \dots, P_k \sim \LLM(\desc)$ with empirical support $\left\vert \support^k(\desc) \right\vert = 1$, i.e. all $k$ samples fall in a single cluster. Then, we can derive a bound on the probability that a future interpretation can be sampled from $\LLM$ that does not fall into the cluster, which would refute semantic collapse. Using $k = \lceil \frac{\log(\delta)}{\log(1 - \epsilon)} \rceil$, it holds that
\begin{equation}
\mathbb{P}( P \not\in \support^k(\desc) \mid P \sim \LLM(\desc)) \leq \epsilon,
\end{equation} with probability of at least $1 - \delta$. In other words, by choosing a sufficiently large $k$, we can bound the risk of incorrectly concluding that the support has collapsed.  

\inlineheadingbf{Sequential Semantic Clustering} Based on this observation, we derive an efficient algorithm for detecting semantic collapse (see \Cref{alg:collapse}). Given a task description $\desc$, we draw samples from $\LLM(\desc)$ one at a time (Line 3), assigning each new program to an existing equivalence class if it is semantically indistinguishable from a representative (Line 4-8), or creating a new class otherwise (Line 10). Sampling terminates early as soon as a second equivalence class is discovered (Line 11) -- refuting collapse -- or once the budget $k$ is exhausted, in which case the empirical support has collapsed to a single interpretation (Line 13). 

This sequential strategy avoids unnecessary sampling: under collapse the full budget is used, but a single divergent sample is sufficient to terminate early.

\inlineheadingbf{Test input generation} The equivalence check in Line 6 of \Cref{alg:collapse} requires a set of representative test inputs $\tests$ to compare program behavior. In general, there are many ways to derive test inputs $\tests$. We follow a common two-stage approach to derive $\tests$ from the task description $\desc$~\cite{DBLP:conf/nips/LiuXW023}: 

\begin{enumerate}[label=\arabic*.]
\item \textbf{Seed Corpus:} An initial set of test inputs is generated by prompting an $\LLM$ with the task description $\desc$, covering representative cases including typical inputs, boundary conditions, and edge cases suggested by the task description. 

\item \textbf{Type-Aware Mutation:} To broaden test diversity beyond what is generated directly, type-aware mutation is applied to the seed corpus. Given a seed input, variants are produced by applying type-preserving perturbations, e.g. boundary values and sign changes for integers, length and content variations for lists, and sub-string changes for strings. This increases the likelihood of exposing behavioral differences between programs that agree on typical inputs but diverge on edge cases.
\end{enumerate}

In our main experiments, $\tests$ is instantiated using inputs from the gold test suite provided with each benchmark, isolating the effect of underspecification on semantic clustering from any confounding effects of automated test generation. The impact of test inputs derived from underspecified task descriptions on semantic clustering is evaluated separately in \Cref{sec:rq2}.

\section{Experimental Setup}

\subsection{Benchmarks}
Our evaluation is based on three widely used code generation benchmarks: \emph{MBPP$^+$}~\cite{DBLP:conf/nips/LiuXW023,DBLP:journals/corr/abs-2108-07732}, \emph{HumanEval$^+$}~\cite{DBLP:conf/nips/LiuXW023,DBLP:journals/corr/abs-2107-03374}, and LiveCodeBench~\cite{DBLP:conf/iclr/JainHGLYZWSSS25}. We consider both the original benchmarks and underspecified variants:

\inlineheadingbf{MBPP$^+$} MBPP$^+$ comprises 378 crowd-sourced Python programming problems, designed to be approachable by entry-level programmers. Each task contains a task description, a reference implementation, and an extensive set of test cases. For MBPP$^+$, we consider three forms of underspecification taken from ~\cite{DBLP:journals/corr/abs-2507-20439}:

\inlineheadingit{Ambiguous} A variant of MBPP$^+$ in which each of the 378 task descriptions is rewritten to introduce ambiguity, admitting multiple plausible interpretations.

\inlineheadingit{Incomplete} A variant of MBPP$^+$ in which each task description omits one or more requirements, leaving aspects of the intended behavior unspecified and admitting multiple valid implementations.

\inlineheadingit{Contradictory} A variant of MBPP$^+$ in which each task description contains inconsistent requirements, forcing a choice between competing resolutions each corresponding to a distinct interpretation.

\inlineheadingbf{HumanEval$^+$} HumanEval$^+$ contains 164 Python programming tasks. The intended behavior is captured by a reference implementation and an extensive set of test cases. For underspecification, we consider a recently introduced dataset of underspecified HumanEval problems~\cite{DBLP:journals/corr/abs-2604-24712}:

\inlineheadingit{Underspecified} A variant of HumanEval$^+$ in which the textual description of the task is underspecified, admitting multiple plausible interepretations.

\inlineheadingbf{LiveCodeBench} LiveCodeBench is a continuously evolving benchmark of programming challenges. The benchmark enables contamination-free evaluation of LLMs by providing challenges published after the knowledge cutoff of LLMs. We employ LiveCodeBench v6 considering 182 problems published after January 1st 2025, which is the knowledge cutoff of all our evaluated models. For underspecification, we adopt a recently introduced underspecified variant~\cite{DBLP:journals/corr/abs-2604-24712}:

\inlineheadingit{Underspecified} A variant of LiveCodeBench in which the task description, e.g. competition-style programming problem, is underspecified, allowing for multiple interpretations.

For all benchmarks, the intended target behavior $P^*$ is defined by the original benchmarks.

\subsection{Evaluation Metrics}
To answer our research questions, we use the following metrics with respect to a sampling budget $k$:

\inlineheadingbf{Pass@k} The likelihood that at least one solution among $k$ sampled LLM responses is correct by passing all test cases of the original benchmark~\cite{DBLP:journals/corr/abs-2107-03374}. 

\inlineheadingbf{Inconsistency ($\mathcal{I}c$)} We measure inconsistency in terms of pointwise incoherence~\cite{DBLP:conf/aaai/ValentinMSB26}. Pointwise incoherence is the likelihood that two random programs (among the $k$ sampled LLM responses for a task $\desc$) can be distinguished by a random test input. 

\inlineheadingbf{Semantic Collapse (\%SC)} The percentage of tasks $\desc$ where $\left\vert \support^k(\desc)\right\vert = 1$. Detrimental semantic collapse is the percentage of tasks that are semantically collapsed tasks with $Pass@k = 0$. 

\subsection{Models}
We consider three state-of-the-art code generation LLMs: Claude Sonnet 4.5~\cite{anthropic2025claude}, GPT-4.1-mini~\cite{DBLP:journals/corr/abs-2303-08774}, Qwen3-32B~\cite{DBLP:journals/corr/abs-2505-09388}. These represent popular coding models: Claude Sonnet 4.5 is one of the most capable coding LLMs~\cite{DBLP:conf/iclr/JainHGLYZWSSS25}, GPT-4.1-mini is an economical option for coding, and Qwen3 32B is an effective open source model that can run on consumer hardware. By default, we sample from the output distribution of these models with a temperature of 0.8 and the full response is evaluated against the test suite of each benchmark. We vary the temperature in the ablation study (\Cref{sec:discussion}).
\section{Results}

\begin{table*}
    \vspace{0.7em}
  \caption{Evaluation results of three representative LLMs under underspecified task descriptions at different sampling budgets $k \in \{1, 5, 10, 25\}$. We report average Pass@k, inconsistency ($\mathcal{I}c$), and semantic collapse (\%SC). Arrows denote the absolute difference in Pass@k to the original benchmark results.   }
      \vspace{0.7em}
  \label{table:results}
  \centering
  \resizebox{\textwidth}{!}{%
  \begin{tabular}{rr rrr r  rrr r  rrr r rrr r rrr}
    \toprule
    && \multicolumn{11}{c}{\bfseries MBPP$^+$} && \multicolumn{3}{c}{\multirow{3}{*}{\shortstack{\bfseries HumanEval$^+$  \\ \bfseries Underspecified}}} && \multicolumn{3}{c}{\multirow{3}{*}{\shortstack{\bfseries LiveCodeBench  \\ \bfseries Underspecified}}} \\
    \cmidrule{3-13} 
    && \multicolumn{3}{c}{\bfseries Ambiguous}   &&   \multicolumn{3}{c}{\bfseries Incomplete}   &&  \multicolumn{3}{c}{\bfseries Contradictory} && \multicolumn{3}{c}{} && \multicolumn{3}{c}{}     \\
      \cmidrule{3-5} \cmidrule{7-9} \cmidrule{11-13}  \cmidrule{15-17} \cmidrule{19-21}
    {\bfseries Model} & $k$ & Pass@$k$ & $\mathcal{I}c$ & \%SC && Pass@$k$ & $\mathcal{I}c$ & \%SC && Pass@$k$ & $\mathcal{I}c$ & \%SC && Pass@$k$ & $\mathcal{I}c$ & \%SC && Pass@$k$ & $\mathcal{I}c$ & \%SC   \\
    \midrule
     & 1 & 44.0\color{red!60}{$\downarrow$-35.7} & 0.0 & 100.0 && 39.1\color{red!60}{$\downarrow$-40.6} & 0.0 & 100.0 && 12.9\color{red!60}{$\downarrow$-66.8} & 0.0 & 100.0 && 87.9\color{red!60}{$\downarrow$ -5.0} & 0.0 & 100.0 && 28.5\color{red!60}{$\downarrow$-2.3} & 0.0 & 100.0\\
     \rowcolor{gray!10}\cellcolor{white}
    & 5 & 49.1\color{red!60}{$\downarrow$-34.3} & 6.5 & 86.5 && 44.2\color{red!60}{$\downarrow$-39.2} & 4.1 & 89.7 && 18.2\color{red!60}{$\downarrow$-65.2} & 7.6 & 84.4 && 91.7\color{red!60}{$\downarrow$ -3.3} & 1.9 & 90.8 && 33.6\color{red!60}{$\downarrow$-3.3} & 20.9 & 54.9 \\
    & 10 & 50.7\color{red!60}{$\downarrow$-33.4} & 15.8 & 59.8 && 45.9\color{red!60}{$\downarrow$-38.2} & 11.9 & 66.1 && 20.1\color{red!60}{$\downarrow$-64.0} & 14.0 & 64.6 && 92.8\color{red!60}{$\downarrow$ -2.7} & 2.7 & 86.6 && 35.6\color{red!60}{$\downarrow$-3.9}& 22.2 & 50.0 \\
    \rowcolor{gray!10}\cellcolor{white}
    \multirow{-4}{*}{\shortstack{Claude \\ Sonnet 4.5}}& 25 & 52.1\color{red!60}{$\downarrow$-33.1} & 15.5 & 52.6 && 47.6\color{red!60}{$\downarrow$-37.6} & 12.5 & 59.5 && 22.0\color{red!60}{$\downarrow$-63.3} & 14.5 & 61.6 && 93.9\color{red!60}{$\downarrow$ -1.8} & 2.9 & 82.3 && 39.0\color{red!60}{$\downarrow$-3.9}& 23.9 & 43.4\\
    \midrule
     & 1 & 36.4\color{red!60}{$\downarrow$-39.4} & 0.0 & 100.0 && 32.7\color{red!60}{$\downarrow$-43.1} & 0.0 & 100.0 && 11.2\color{red!60}{$\downarrow$-64.6} & 0.0 & 100.0 && 71.4\color{red!60}{$\downarrow$-12.7} & 0.0 & 100.0 && 25.2\color{red!60}{$\downarrow$-2.9} & 0.0 & 100.0 \\
     \rowcolor{gray!10}\cellcolor{white}
    & 5 & 43.5\color{red!60}{$\downarrow$-37.1} & 10.2 & 79.4 && 38.9\color{red!60}{$\downarrow$-41.7} & 7.9 & 81.0 && 15.9\color{red!60}{$\downarrow$-64.7} & 8.7 & 80.7 && 75.9\color{red!60}{$\downarrow$-15.3} & 2.4 & 89.6 && 30.9\color{red!60}{$\downarrow$-4.1}  & 23.4 & 55.5 \\
    & 10 & 45.8\color{red!60}{$\downarrow$-35.9} & 20.8 & 50.3 && 41.4\color{red!60}{$\downarrow$-40.3} & 17.4 & 56.1 && 17.9\color{red!60}{$\downarrow$-63.8} & 18.2 & 57.1 && 78.0\color{red!60}{$\downarrow$-14.4} & 6.5 & 76.2 && 33.5\color{red!60}{$\downarrow$-4.8} & 23.7 & 53.3 \\
    \rowcolor{gray!10}\cellcolor{white}
    \multirow{-4}{*}{\shortstack{GPT-4.1\\mini}}& 25 & 48.4\color{red!60}{$\downarrow$-34.4} & 21.7 & 43.9 && 44.7\color{red!60}{$\downarrow$-38.1} & 17.6 & 48.4 && 20.3\color{red!60}{$\downarrow$-61.4} & 19.0 & 50.3 && 80.5\color{red!60}{$\downarrow$-12.8} & 6.9 & 70.1 && 36.8\color{red!60}{$\downarrow$-5.7} & 24.5 & 48.4\\
    \midrule
     & 1 & 39.0\color{red!60}{$\downarrow$-37.5} & 0.0 & 100.0 && 32.7\color{red!60}{$\downarrow$-43.8} & 0.0 & 100.0 && 11.1\color{red!60}{$\downarrow$-65.4} & 0.0 & 100.0 && 61.2\color{red!60}{$\downarrow$ -6.8} & 0.0 & 100.0 && 34.4\color{red!60}{$\downarrow$-2.6} & 0.0 & 100.0  \\
     \rowcolor{gray!10}\cellcolor{white}
    & 5 & 48.6\color{red!60}{$\downarrow$-36.3} & 7.3 & 85.7 && 38.9\color{red!60}{$\downarrow$-46.0} & 10.2 & 79.1 && 15.9\color{red!60}{$\downarrow$-69.0} & 8.5 & 80.7 && 67.9\color{red!60}{$\downarrow$ -4.8} & 2.6 & 92.7 && 44.6\color{red!60}{$\downarrow$-3.8} & 28.9 & 48.4  \\
    & 10 & 51.2\color{red!60}{$\downarrow$-35.2} & 9.9 & 73.0 && 41.4\color{red!60}{$\downarrow$-45.0} & 21.0 & 50.3 && 17.9\color{red!60}{$\downarrow$-68.5} & 18.1 & 57.4 && 69.1\color{red!60}{$\downarrow$ -4.5} & 3.4 & 89.6 && 48.0\color{red!60}{$\downarrow$-4.4} & 29.2 & 44.5 \\
    \rowcolor{gray!10}\cellcolor{white}
    \multirow{-4}{*}{\shortstack{Qwen3 \\ 32B}}& 25  & 54.8\color{red!60}{$\downarrow$-33.3} & 17.5 & 23.8 && 44.7\color{red!60}{$\downarrow$-43.4} & 17.7 & 43.4 && 20.4\color{red!60}{$\downarrow$-67.7} & 18.6 & 50.0 && 70.1\color{red!60}{$\downarrow$ -4.3} & 11.6 & 42.7 && 52.2\color{red!60}{$\downarrow$-3.8} & 29.4 & 43.4\\

    \bottomrule
  \end{tabular}
  }
      \vspace{0.7em}
\end{table*}

\begin{figure}[t]
    \vspace{0.6em}
\centering
\includegraphics[width=1\linewidth]{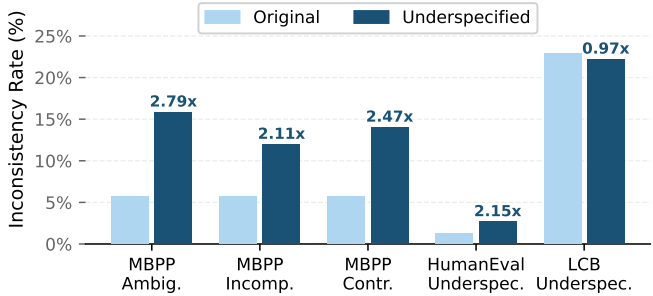}
    \vspace{0.5em}
\caption{\textbf{Expected: Underspecification increases inconsistency {\em on average}.} Relative increase of inconsistency for Claude Sonnet 4.5.} 
    \vspace{0.6em}
\label{fig:inconsistency}
\end{figure}

\subsection{RQ1 - Underspecification and Diversity}
To answer RQ1, we start to evaluate the impact of prompt underspecification on the diversity of LLM responses for sampling budgets $k \in \{1, 5, 10, 25\}$. We use $k = 10$ as our default configuration.
\Cref{table:results} shows our experimental results, reporting metrics on underspecified benchmarks alongside absolute difference to the original benchmarks. 

\inlineheadingbf{Underspecification increases inconsistency \emph{on average}} Increasing the sampling budget from $k = 1$ to $k = 25$ increases the Pass@k of all LLMs, suggesting that LLMs explore different interpretations of the task description and eventually discover the correct interpretation. Compared to the original benchmark (see \Cref{fig:inconsistency}), inconsistency increases by a factor of 2.79x, confirming that LLM responses on underspecified tasks are more diverse {\em on average}. 

\inlineheadingbf{LLMs often miss the correct interpretation under underspecification} While increased inconsistency is expected and has motivated previous works, we observe a significant gap in Pass@k of up to 66.8\% compared to the original benchmark. For MBPP, this gap persists at higher sampling budgets (e.g. up to Pass@25), suggesting that repeated sampling alone is insufficient for LLMs to reliably produce the correct interpretation of underspecified tasks. On HumanEval and LiveCodeBench, where task descriptions include example input-output pairs that partially compensate for underspecification~\cite{DBLP:journals/corr/abs-2604-24712}, the gap is substantially smaller but still persists.

\inlineheadingbf{Semantic collapse persists for underspecification} Surprisingly,  semantic collapse persists for a large fraction of tasks from $k = 5$ to $k = 25$. At $k = 10$, between $50.3$\% to $73.0$\% of underspecified MBPP tasks, varying across benchmarks and models, produce semantically indistinguishable programs across all samples.
Increasing the sampling budget to $k = 25$ reduces collapse, yet it sill persists for $23.8\%$ to $61.6$\% of tasks. Consistent trends are observed for HumanEval and LiveCodeBench. 

This means that while inconsistency increases on average, LLMs still converge to a single interpretation for a large fraction of underspecified tasks, providing no signal to clustering-based approaches that the task description was underspecified.

\llbox{\emph{Underspecification does not imply inconsistency.} While underspecification increases inconsistency on average, suggesting that LLMs explore multiple inconsistent interpretations, this aggregated trend masks a per-task reality: LLMs frequently collapse to a single interpretation on a large fraction of underspecified tasks, providing no signal that the description was underspecified.}

\subsection{RQ2 - Detrimental Semantic Collapse}\label{sec:rq2}

\begin{figure}[t]
\centering
\includegraphics[width=0.99\linewidth]{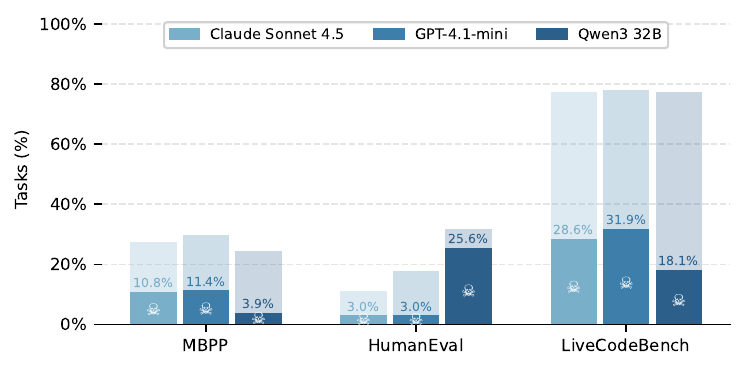}
\caption{\textbf{Unexpected: 1 in 10 MBPP tasks show detrimental collapse.} Percentage of tasks with detrimental semantic collapse. The light color show the percentage of tasks with erroneous responses.} 
    \vspace{0.5em}
\label{fig:detrimental}
\end{figure}

Not all semantic collapse is \emph{detrimental}: converging on the correct interpretation is often desirable. To answer RQ2, we measure the frequency of detrimental collapse and whether prompt underspecification increases its rate relative to the original benchmark. Overall, we find that:

\inlineheadingbf{Detrimental collapse is pervasive across benchmarks}  \Cref{fig:detrimental} shows the rate of detrimental collapse. Detrimental collapse is non-trivial across all original descriptions of the studied benchmarks but varies considerably: over 3\% of task on HumanEval, 10 to 16\% on MBPP, and 18\% to 32\% on LiveCodeBench exhibit detrimental collapse. 

The substantially higher rate on LiveCodeBench reflects the interplay between task difficulty and semantic collapse: competitive programming tasks have inherently low pass rate, yet LLMs still confidently produce semantically indistinguishable programs, frequently converging on an incorrect interpretation with no indication of uncertainty.

\begin{figure*}[t]
\vspace{0.5em}
\centering
\centering
\begin{subfigure}{0.32\linewidth}
    \includegraphics[width=\linewidth]{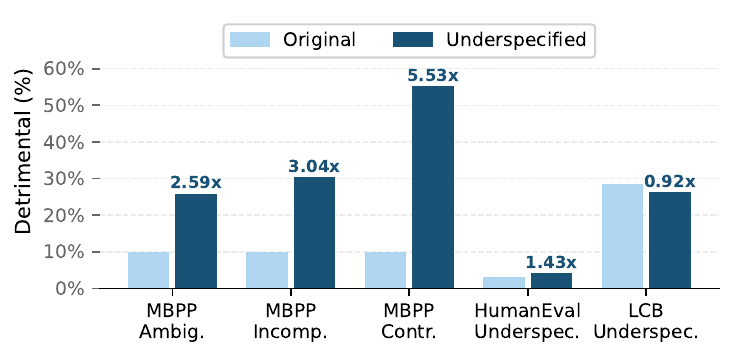}
    \caption{Claude Sonnet 4.5}\label{fig:mbpp}
\end{subfigure}
\hfill
\begin{subfigure}{0.32\linewidth}
     \includegraphics[width=\linewidth]{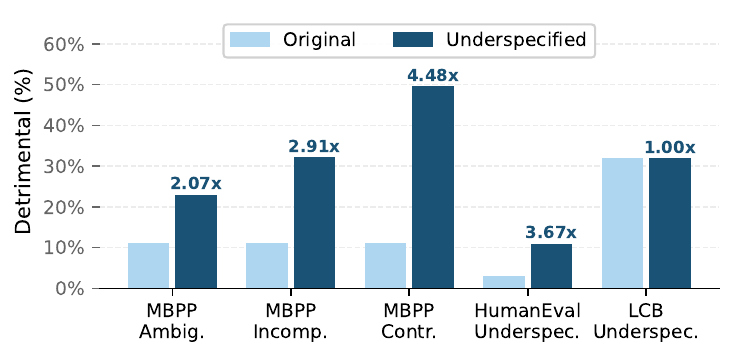}
    \caption{GPT-4.1-mini}\label{fig:clustering}
\end{subfigure}
\hfill
\begin{subfigure}{0.32\linewidth}
    \includegraphics[width=\linewidth]{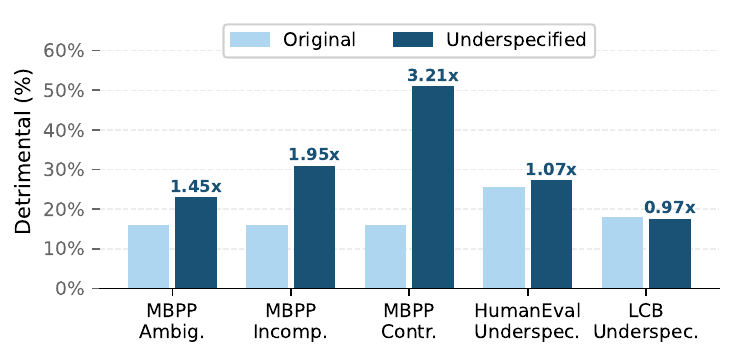}
    \caption{Qwen3 32B}\label{fig:collapse}
\end{subfigure}
\vspace{0.5em}
\caption{\textbf{Detrimental collapse increases with underspecification.} Percentage of tasks with detrimental collapse. Bold numbers indicate the relative increase of detrimental collapse between the original benchmark and the underspecified variant. } 
\label{fig:detrimental-underspec}
\vspace{0.5em}
\end{figure*}

\inlineheadingbf{Underspecification increases detrimental collapse} \Cref{fig:detrimental-underspec} shows the increase in detrimental semantic collapse on underspecified task descriptions relative to the original benchmark. For MBPP, detrimental collapse increases by a factor of $1.45\times$ to $5.53\times$ depending on the underspecification type and model, making it the dominant failure mode: between 23\% and 55\% of underspecified tasks exhibit detrimental collapse, compared to 10 -- 11\% on the original task descriptions. For HumanEval, underspecification increases detrimental collapse by a factor of 1.07x to $3.67\times$. On LiveCodeBench, underspecification has only a marginal impact of a factor of $0.92\times$ to $1.00\times$.

\begin{figure}[t]
\centering
    \vspace{0.5em}
\includegraphics[width=0.99\columnwidth]{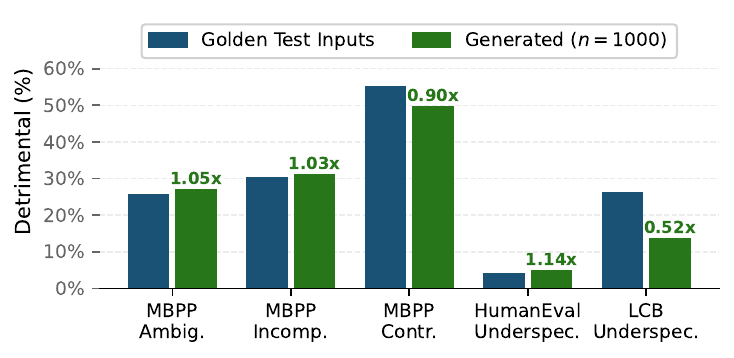}
\vspace{0.5em}
\caption{\textbf{Impact of LLM-based test generation.} Percentage of tasks with detrimental collapse for Claude Sonnet 4.5 as decided by the golden test inputs and generated test inputs ($n = 1000$). Bold numbers indicate the relative difference in detrimental collapse. Decrease stems from the test inputs being better aligned with the LLM's understanding of the task.} 
\label{fig:testgen}
\vspace{0.5em}
\end{figure}

\inlineheadingbf{Impact of LLM-generated test inputs}
Semantic clustering is in practice often performed by executing programs on
LLM-generated test inputs~\cite{DBLP:journals/pacmse/Mu00YZWL024, DBLP:conf/aaai/ValentinMSB26}. We generate $n = 1{,}000$ test inputs (see \Cref{sec:measuring}) and
use them to partition programs into semantic clusters. \Cref{fig:testgen} shows the
impact of LLM-generated test inputs on detrimental collapse for Claude Sonnet
4.5 on the underspecified benchmarks; results for GPT-4.1-mini and Qwen3 32B
follow the same trend. LLM-generated test inputs increase the rate of
detrimental collapse on most benchmarks, with multipliers ranging from
$1.03\times$ (MBPP Incomp.) to $1.14\times$ (HumanEval Underspec.), suggesting
that generated tests are less effective at distinguishing between incorrect
solutions. A notable exception is LiveCodeBench Underspec., where detrimental
collapse decreases by a factor of $0.52\times$, which we attribute to generated
test inputs being better aligned with the LLM's understanding of the task. Across all variants, detrimental collapse persists under
LLM-generated tests, confirming that detrimental collapse persists under LLM-generated tests. 

\llbox{\emph{Incorrectness does not imply inconsistency.} Semantic collapse becomes detrimental in 3\% to 32\% of tasks of the original benchmark set, increasing by a factor of up to  5.5x under prompt underspecification. Detrimental collapse accounts for a substantial fraction of incorrectly solved tasks, providing no indication of incorrectness.}

\subsection{RQ3 - Impact on Disambiguation}

To answer RQ3, we follow existing work~\cite{DBLP:journals/pacmse/Mu00YZWL024} to simulate clarification dialogues in the presence of prompt underspecification. 

\inlineheadingbf{Disambiguation simulation} We simulate a multi-turn clarification conversation following the setup of ClarifyGPT~\cite{DBLP:journals/pacmse/Mu00YZWL024}, replacing the human user with an LLM prompted with the ground truth intent $P^*$. The simulation process proceeds as follows: (1) The user initiates the conversation with a task description $\desc$; (2) the LLM samples up to $k$ programs from $\LLM(\desc)$; (3) if multiple clusters are produced, the LLM generates a clarifying question which the user answers based on $P^*$; the task description is updated, and the process repeats from step (2). (4) If all samples fall into a single cluster, a solution is returned from that cluster and the conversation ends.

\begin{figure}[t]
\centering
    \vspace{0.5em}
    \includegraphics[width=\columnwidth]{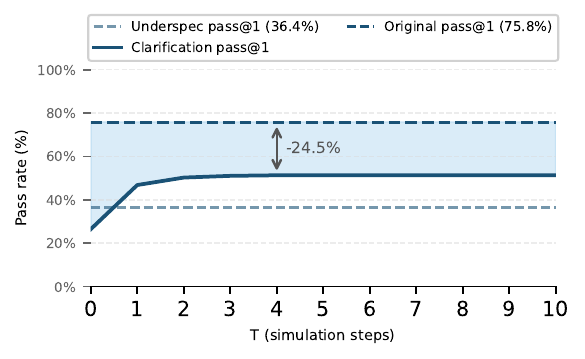}
        \vspace{0.5em}
\caption{\textbf{Conversation simulation on ambiguous MBPP tasks.} The blue line shows the pass rate of GPT-4.1-mini generated solutions after asking $T$ questions. The light dashed line is the baseline performance without asking questions. The darker dashed line shows the performance of GPT-4.1-mini on the original benchmark.  } 
    \vspace{0.5em}
\label{fig:simulation}
\end{figure}

\inlineheadingbf{Disambiguation improves performance by decreasing inconsistency}
\Cref{fig:simulation} illustrates the conversation behaviour of GPT-4.1-mini
on MBPP-Ambiguous. By asking clarifying questions, disambiguation improves
code generation performance from 36.4\% to 46.8\% after the first turn, and
further to 51.3\% after around 4 turns. \Cref{table:disambiguation} summarises the conversation simulations across all benchmarks. Disambiguation produces consistent improvements over the underspecified baseline: from 36.4\% to 51.3\% on MBPP-Ambiguous, from 32.7\% to 51.3\% on MBPP-Incomplete, from
11.2\% to 13.8\% on MBPP-Contradictory, from 71.4\% to 82.9\% on HumanEval-Underspec., and from 25.2\% to 26.4\% on LiveCodeBench-Underspec.

These improvements are driven by a reduction in inconsistency: conversation simulations converge to a single semantic cluster within 0 to 9 turns, indicating that clarification is an effective means of achieving consistency.

\inlineheadingbf{Disambiguation leaves a substantial gap to the original benchmark performance}
Despite the improvements reported above, \Cref{fig:simulation} and
\Cref{table:disambiguation} reveal a substantial remaining gap to the original
benchmark performance. On MBPP in particular, disambiguated performance remains
24.5\% to 62.0\% below the original benchmark even after full clarification.
The gap is considerably smaller for HumanEval ($-1.2\%$) and LiveCodeBench ($-1.7\%$), suggesting that the degree of underspecification in MBPP is fundamentally harder to resolve through clarification alone. 

This highlights that clustering-based disambiguation, while effective at reducing inconsistency,
does not fully resolve underspecification.

\begin{table}
  \caption{Summary of conversation simulations across underspecified benchmarks. We report the final pass rate, the minimum -- maximum number of turns, the detrimental semantic collapse at turn 0 (DC@0), after one clarification question (DC@1),  after follow-up questions (DC@10). Arrows denote the absolute difference of pass rate and Pass@1 of the original benchmark tasks.}
  \vspace{0.5em}
  \label{table:disambiguation}
  \centering
  \resizebox{\columnwidth}{!}{%
  \begin{tabular}{lr rr | r r  rrr r  rrr r  rrr r rrr}
    \toprule
     \textbf{Benchmark} && \textbf{Pass rate} & \textbf{Turns}  & \textbf{DC@0} & \textbf{DC@1} & \textbf{DC@10} \\
    \midrule
    MBPP-Ambiguous &&  51.3\color{red!60}{$\downarrow$-24.5}  & 0 -- 6 & 23.0 & 37.3 & 48.7\\
    MBPP-Incomplete &&  51.3\color{red!60}{$\downarrow$-24.5} & 0 -- 5 & 32.3 &  42.1 & 48.7\\
    MBPP-Contradictory &&  13.8\color{red!60}{$\downarrow$-62.0} & 0 -- 7 & 49.7 &  76.5 & 86.2\\
    \midrule
    HumanEval-Underspec. && 82.9\color{red!60}{$\downarrow$ -1.2} & 0 -- 4 & 11.0 &  13.4 & 17.1 \\
    LCB-Underspec. && 26.4\color{red!60}{$\downarrow$ -1.7} & 0 -- 9 & 31.9 & 48.4 & 73.6 \\
    \bottomrule
  \end{tabular}
  }
\end{table}

\inlineheadingbf{Detrimental collapse is a blind spot}
As shown in \Cref{table:disambiguation}, detrimental semantic collapse is a
major blind spot of clustering-based disambiguation: from 11\% to 49.7\% of
tasks receive an incorrect solution without triggering a clarifying question.
Moreover, asking questions does not always produce correct results: after one clarification turn, detrimental collapse increases to 37.3\%--76.5\% on MBPP variants. This is in part because clarifying questions are grounded in the behavioural differences the model itself produces: if two incorrect solutions for example differ only in edge case handling, the clarifying question will concern edge cases rather than the more fundamental source of incorrectness.

\llbox{\emph{Detrimental semantic collapse is a fundamental blind spot of clustering-based disambiguation.} Even when a simulated user provides clarifications, disambiguation fails silently on tasks affected by detrimental collapse: the model converges to a single incorrect interpretation without ever asking a question, returning a wrong solution with full confidence.}

\section{Discussion}
\label{sec:discussion}
We now discuss the risks of detrimental collapse in coding models and its implications for existing and future research studies that rely on semantic clustering.  

\inlineheadingbf{Model uncertainty and inconsistency}
Our results indicate that neither underspecification nor incorrectness implies
inconsistency. We therefore argue that inconsistency should be interpreted for
what it is: an \emph{expression of model uncertainty}. 
Prompt underspecification and misunderstanding can each induce model
uncertainty, and we find in RQ1 that underspecification does indeed increase average
inconsistency. Yet our results also reveal that a large fraction of tasks
exhibit detrimental collapse: the model remains highly certain yet ultimately
fails to solve the task correctly.

\inlineheadingit{Risks for clustering-based methods} A core consequence of our findings is that any semantic clustering-based
detection method has a \emph{blind spot} dependent on the LLM and benchmark.
For disambiguation methods, detrimental collapse risks returning a confidently
incorrect solution for an underspecified task without prompting for
clarification. For correctness estimation, it causes models that are
confidently incorrect to be systematically overvalued.

\inlineheadingbf{Ablation}
Semantic clustering is sensitive to the sampling budget $k$ and the sampling temperature $\tau$. We evaluate the impact of both parameters on detrimental collapse. Our results are shown in  \Cref{table:ablation}.

\inlineheadingit{Sampling budget} As the cost of frontier LLMs has increased, sampling budgets in existing work
have decreased over time: \cite{DBLP:journals/pacmse/Mu00YZWL024} evaluated GPT-3.5 at $k = 25$, \cite{DBLP:conf/aaai/ValentinMSB26} employed $k = 10$ for more recent models, and \cite{DBLP:journals/corr/abs-2502-11620} evaluated at $k = 5$. Our ablation (\Cref{table:ablation}) confirms that larger $k$ reduces detrimental collapse, and that this trend needs to be reversed to mitigate detrimental collapse. We use $k = 10$ as our default as a practical
compromise: increasing from $k = 10$ to $k = 25$ reduces detrimental collapse by 2.6 percentage points on MBPP and 6.0 percentage points on LiveCodeBench,
but at 2.5$\times$ the cost. In practical settings, this trade-off is likely to favor lower $k$ values, given that the incurred cost is proportional to all coding prompting tasks, even those involving fault fixing. 

\inlineheadingit{Temperature}
The sampling temperature controls the determinism of an LLM's output. A temperature close to zero yields near-deterministic outputs, while a temperature close to or above one often deteriorates performance~\cite{DBLP:journals/corr/abs-2107-03374}. Our ablation (\Cref{table:ablation}) shows that increasing temperature consistently reduces detrimental collapse, from 15.1\% to 11.1\% on MBPP and from 38.5\% to 32.4\%
on LiveCodeBench at no additional cost. We use $\tau = 0.8$ as our default, as it is commonly adopted in code generation benchmarks to maximise performance~\cite{DBLP:conf/iclr/JainHGLYZWSSS25}.

\inlineheadingit{Practical implications} Detrimental collapse can be partly mitigated by increasing the sampling budget, but at a direct cost to practitioners. At the time of writing, solving 378 MBPP tasks with Claude Sonnet 4.5 costs approximately \$1.30 at $k = 1$, \$6.52 at $k = 5$, \$13.04 at $k = 10$, and \$32.60 at $k = 25$. Reducing the risk of missing an interpretation to below 1\% (at $\delta = 95\%$ confidence;
see \Cref{sec:measuring}) would require $k = 300$, incurring an estimated cost of over \$391 per benchmark run. This cost becomes prohibitive in large-scale deployment, where semantic clustering would need to be applied across thousands
of functions.

\begin{table}
\vspace{0.5em}
  \caption{Ablation analysis of sampling parameters on the detrimental collapse of GPT-4.1-mini. We report percentage of task with detrimental collapse (\%DC) and the cost of the performed evaluation (\$).}
  \vspace{0.5em}
  \label{table:ablation}
  \centering
  \resizebox{\columnwidth}{!}{%
  \begin{tabular}{lr rrr r  rrr r  rrr r  rrr r rrr}
    \toprule
     & \multicolumn{2}{c}{\textbf{MBPP}$^+$} && \multicolumn{2}{c}{\textbf{HumanEval}$^+$}  && \multicolumn{2}{c}{\textbf{LiveCodeBench}} \\
     \cmidrule{2-3} \cmidrule{5-6} \cmidrule{8-9}
     & \%DC & \$ && \%DC & \$ && \%DC & \$ \\
    \midrule
    \rowcolor{gray!10}
    \multicolumn{9}{c}{Detrimental collapse at different $k$}\\
    \midrule
    $k = 1$ & 24.1 & \$0.14 && 12.1 & \$0.08 && 73.1 & \$0.20 \\
    $k = 5$ & 19.6 & \$0.72 && 6.7 & \$0.41 && 52.2 & \$1.02  \\
    $k = 10$ & 11.1 & \$1.44 && 3.0 & \$0.82 && 32.4 & \$2.04 \\
    $k = 25$ & 8.5 & \$3.59 && 3.0 & \$2.05 && 26.4 & \$5.10 \\
    \midrule
    \rowcolor{gray!10}
    \multicolumn{9}{c}{Detrimental collapse at different temperature $\tau$}\\
    \midrule
    $\tau = 0.2$ & 15.1 & \$1.44 && 6.1 & \$0.82 && 38.5 & \$2.04\\
    $\tau = 0.6$ & 11.1 & \$1.44 && 3.0 & \$0.82 && 37.4 & \$2.04\\
    $\tau = 0.8$ & 11.1 & \$1.44 && 3.0 & \$0.82 && 32.4 & \$2.04 \\
    $\tau = 1.0$ & 10.3 & \$1.44 && 3.0 & \$0.82 && 30.1 & \$2.04 \\
    \bottomrule
  \end{tabular}
  }
  \vspace{0.5em}
\end{table}

\section{Threats to Validity}

\subsection{External validity} Our findings are derived from Python code generation tasks which might thus limit the generalization of our finding. We performed our experiments on three popular programming benchmarks (MBPP, HumanEval, and LiveCodeBench). MBPP and HumanEval focus on relatively small self-contained functions, while LiveCodeBench is a collection of competition-level programming tasks. Although these tasks are varied in complexity and topic, they do not cover all possible coding scenarios. We consider that our findings will be amplified in more complex projects, where cost of sampling becomes an even greater concern. 

The underspecification in our benchmarks is artificially introduced~\cite{DBLP:journals/corr/abs-2507-20439}. While validated by experts, the types of underspecification might not reflect all forms of underspecification observed in the real world. Furthermore, the concrete instantiation of underspecification might not be representative for real-world underspecification. We however observe that these specification flaws exists in real-world specification and can induce detrimental collapse. 

We evaluated three popular state-of-the-art coding models, namely Claude Sonnet 4.5, GPT-4.1-mini, and Qwen3 (32B). This diversity of models and tasks increases confidence that our conclusions are not tied to a specific model or dataset.

\subsection{Internal validity} Our methodology to detect semantic collapse is sensitive to the sampling budget $k$. Since we limit the budget to $k \in \{1, 5, 10, 25\}$, our method might have prematurely concluded semantic collapse. As these sampling rates are commonly used by existing methods, our results remain practically relevant and represent a true failure mode of semantic clustering-based approaches. 
We further employ LLMs to simulate user responses during disambiguation. Real user responses may deviate from simulated ones in practice, and hence our results may not generalize to real user interactions. 

Nevertheless,
detrimental collapse will cause clustering-based disambiguation to withhold clarifying questions regardless of the user simulation, as it occurs before
any interaction takes place.

A potential threat is that we sample responses from the LLM provider's API, which might implement caching mechanisms that reduce output variability. We mitigate this threat by sampling model responses across different sessions and varying the API keys used. In addition, we observe detrimental collapse for both closed- and open-source models, indicating that detrimental collapse is a fundamental property of LLMs rather than an artifact of provider-specific caching mechanisms.

Another potential confounding factor is memorization. The tasks may have been seen during the training, inflating the performance of the LLMs. In this case, detrimental collapse might represent a failure mode of the training process, i.e. the model collapses because of a wrong training signal. However, our findings generalize to a contamination-free subset of LiveCodeBench, indicating that the failure mode persists beyond what have been seen during training.

\subsection{Construct validity} 

Construct validity concerns the extent to which operationalizations used in the study accurately capture the underlying concepts of interest—namely prompt underspecification, model incoherence, and detrimental semantic collapse. 

Incoherence is measured via semantic clustering of generated programs, grouping implementations according to their observed execution behavior on test cases. This operationalization is inherently dependent on the quality of the employed test suites. If test cases are incomplete or fail to distinguish semantically different implementations, distinct behaviors may be grouped into the same cluster, leading to an underestimation of incoherence. Conversely, overly sensitive distinctions in clustering may exaggerate differences that are not semantically meaningful. Thus, the mapping from behavioral equivalence to coherence is only approximate.

Coherence is inferred from output consistency across repeated generations, but such consistency may arise from factors unrelated to semantic understanding. For instance, the use of decoding strategies (e.g., temperature settings), training data biases, or model priors may experience low semantic collapse. Additionally, the analysis is based solely on observable outputs, without access to the model’s internal reasoning processes. As a result, the study cannot distinguish between true semantic collapse—where the model fails to represent alternative interpretations—and cases where multiple interpretations are internally considered but not expressed in the final output.

\section{Conclusion}
Semantic clustering, i.e. grouping LLM-generated programs by their observable
behaviour, is increasingly applied to detect prompt underspecification and model incorrectness in coding tasks. As both are critical challenges in code
generation, understanding the capabilities and limitations of clustering-based approaches is essential.

In this work, we studied the inconsistency of three state-of-the-art coding models under underspecification, and identified {\em detrimental semantic collapse} as a critical blind spots. Our results show that models collapse detrimentally on a substantial fraction
of tasks, ranging from 3\% in HumanEval to over 32\% in LiveCodeBench, and that underspecification significantly increases the rate of detrimental
collapse. Crucially, detrimental collapse is not an artifact of correct solutions as previously assumed: models frequently commit to a single incorrect interpretation even when they fail to solve the task, rendering inconsistency-based detection ineffective precisely where it is needed most.

These findings have direct implications for methods that rely on semantic clustering as a proxy for incorrectness or underspecification. Our results show that inconsistency
can be a meaningful signal when present, but its absence cannot be interpreted as evidence of correct specification or solution. 

We further demonstrate that the sampling budgets prevalent in the literature are likely insufficient to reliably distinguish collapse from genuine consistency, a gap that cannot be closed by simply increasing $k$, given the fundamental statistical limits of finite-sample collapse detection. The practical implication is that developers must spend a significant share
of their budget on resampling and evaluating code candidates to apply clustering-based methods, without gaining substantially more \emph{trust} in their correctness.

We hope these findings prompt the community to revisit the assumptions underlying clustering-based evaluation and disambiguation methods, and to develop detection approaches that remain effective even in the absence of observable inconsistency. This could be for instance achieved by leveraging internal uncertainty signals, reasoning traces, and diversity-aware sampling strategies that may better explore alternative semantic interpretations. Further work is also needed to understand the underlying causes of collapse, establish theoretical limits for finite-sample collapse detection, and develop benchmarks with controlled levels of underspecification. 

\IEEEtriggeratref{81} 
\bibliographystyle{IEEEtranS}
\bibliography{literature}

\end{document}